# Interaction of quantum dot molecules with multi-mode radiation fields

**Amir Hossein Sadeghi, Ali Naqavi, Sina Khorasani**

*School of Electrical Engineering, Sharif University of Technology*

*P. O. Box 11365-9363, Tehran, Iran*

**Correspondence:**

Sina Khorasani

*Associate Professor of Electrical Engineering*

School of Electrical Engineering, Sharif University of Technology,

Azadi Ave., Tehran, P. O. Box 11365-9363, Iran

Tel.: +98-21-6616-4352

Fax: +98-21-6602-3261

E-mail: khorasani@sharif.edu

## ABSTRACT

In this article, the interaction of an arbitrary number of quantum dots, behaving as artificial molecules, with different energy levels and multi-mode electromagnetic field is studied. We make the assumption that each quantum dot can be represented as an atom with zero kinetic energy, and that all excitonic effects except dipole-dipole interactions may be disregarded. We use Jaynes-Cummings-Paul model with applications to quantum systems based on a time-dependent Hamiltonian and entangled states. We obtain a system of equations describing the interaction and present a method to solve the equations analytically for a single mode field within the Rotating-Wave Approximation. As an example of the applicability of this approach, we solve the system of two two-level quantum dots in a lossless cavity with two modes of electromagnetic field. We furthermore study the evolution of entanglement by defining and computing the concurrency.

## INTRODUCTION

Quantum entangled states have become a popular topic for research in the last decade due to their potential application in quantum communication and information [1-3]. Generally, the trend of research is towards increasing the number of entangled bodies and to our knowledge, entanglement of eight particles has been the state of the art [4]. Parallel to experiments, theoretical work has also been devoted to modeling phenomena associated with the entanglement [5-10]. The simplest model for description of interaction within the framework of Quantum Electrodynamics (QED) is the so-called Jaynes-Cummings-Paul Model (JCPM) [11], which provides a closed-form and explicit solution to the case of a two-level atom with zero kinetic energy in a cavity with a single mode electromagnetic field. In real world, however, the problem is much more complex: for instance, neither the field is necessarily single mode nor do the atoms possess only two energy levels. Also, quantum dots with various configurations may be exploited as artificial atoms; combinations of quantum dots give rise to the concept of quantum dot (artificial) molecules. Hence, more sophisticated models have been proposed to include multi-atom systems [12], multi-phonon transitions [13], intensity-dependent entanglement [14], three and more energy levels [15] and electromagnetic modes [15, 16].

In this paper, we provide a generalization of the JCPM to investigate a system of arbitrary number of quantum dots (or atoms with negligible kinetic energy), with arbitrary number of energy levels in a multi-mode electromagnetic (EM) field. First, we will find an appropriate Hamiltonian for the system. Such a Hamiltonian should account for dot-dot and dot-field interaction while being simple enough to avoid increasing the solution complexity. We employ the Heisenberg's interaction picture [17] to obtain such a Hamiltonian, here being referred to as the image Hamiltonian. We devise a rigorous and algorithmic method of substituting the image Hamiltonian into the time dependent Schrödinger's equation, to construct a system of simultaneous differential equations. This system of linear ordinary differential equations are regarded as a generalization of well-known Rabi equations in the JCPM. By solving this system of equations, we find the time evolution of the system and therefore its entanglement, through evaluation of a proper functional, which is referred to as the concurrency. We ignore spin and excitonic effects in quantum dots to simplify the formulation, however, the latter effect has been thoroughly discussed in a recent paper of our group [18].

Finding a solution for the mentioned system of equations analytically seems to be a sophisticated task in the general case. Actually, further assumptions such as the Rotating Wave Approximation (RWA) can greatly help one to simplify the problem. We show that for single-mode fields, the RWA leads to an analytically solvable problem, however, finding an analytical solution is at least much more complex for multi-mode EM fields. In such cases, numerical methods can replace analytical methods. We present examples for both cases of single- and multi-mode EM fields. In the former case, we follow the analytical approach while for the latter case, we have implemented a simple finite difference scheme to calculate the time evolution of the system. It is furthermore shown that, if the EM field is not single mode, the system should be described through a system of differential equations with the number of equations dependent on the photon numbers of the EM modes. Unfortunately, the number of equations blows up rapidly as the number of EM modes increases and this, imposes an undesirable computational burden.

Evaluating the degree of entanglement is the final step in this work. Several works have been previously performed to provide an appropriate tool to measure multi-particle entanglement [3,19,20] among which, we have used the generalized definition of concurrency as mentioned in [21] to present entanglement in the last example- two two-level dots in a cavity with two EM modes.

## BASIC DEFINITIONS

In this article, the interaction of a collection of $k$-dots with arbitrary energy levels and a reservoir of $w$-modes of electromagnetic field is investigated. Each dot interacts with light through specified constants according to the following Hamiltonian [17]

$$\mathbb{H}_{r.E} = \sum_{n\ ,i<j} \left(\gamma_{nij}\hat{\sigma}^n_{i,j} + \gamma^*_{nij}\hat{\sigma}^n_{j,i}\right) \sum_{\nu} \left(g_{nij\nu} a_\nu + g^*_{nij\nu} a^\dagger_\nu\right), \tag{1}$$

where coefficients $\gamma_{nij}$ are matrix elements of the dipole operator of dot indexed by $n$ and coefficient $g_{nij\nu}$ determines the coupling strength of dot $n$ with the field of mode indexed by $\nu$. Both of $\gamma_{nij}$ and $g_{nij\nu}$ are assumed to be dependent on the two energy levels $i$ and $j$ between which, the electron of dot $n$ do a transition through the interaction with the field. Operator $\hat{\sigma}^n_{i,j}$ is the transition operator from level $j$ into level $i$ for dot $n$ and $a_\nu$ and $a^\dagger_\nu$ are annihilation and creation operators of the photon corresponding to mode $\nu$ respectively. By indicating the $i$-th eigenket of dot $n$ as $\left|\begin{matrix} n \\ i \end{matrix}\right\rangle$, we may adopt the following definitions of atomic operators

$$\hat{\sigma}^n_{i,j} = \left|\begin{matrix} n \\ i \end{matrix}\right\rangle \left\langle\begin{matrix} n \\ j \end{matrix}\right|, \tag{2}$$

$$\hat{\sigma}^n_i \triangleq \hat{\sigma}^n_{i,i} = \hat{\sigma}^n_{i,j} = \left|\begin{matrix} n \\ i \end{matrix}\right\rangle \left\langle\begin{matrix} n \\ i \end{matrix}\right|, \tag{3}$$

$$\hat{\sigma}^n_{i,j}\hat{\sigma}^n_{k,i'} = \hat{\sigma}^n_{i,i'}\delta_{j,k}, \tag{4}$$

$$\hat{\sigma}^n_{i,j}\left|\begin{matrix} n \\ k \end{matrix}\right\rangle = \left|\begin{matrix} n \\ i \end{matrix}\right\rangle \left\langle\begin{matrix} n \\ j \end{matrix}\right|\left|\begin{matrix} n \\ k \end{matrix}\right\rangle = \left|\begin{matrix} n \\ i \end{matrix}\right\rangle \delta_{j,k}, \tag{5}$$

$$\hat{\sigma}^s_f\hat{\sigma}^s_g = \delta_{f,g}\hat{\sigma}^s_f, \tag{6}$$

$$\sum_f \hat{\sigma}^s_f = 1. \tag{7}$$

where $\delta_{f,g}$ represents the Kronecker delta function. Relation (6) is a result of orthogonality of the eigenstates and (7) is correct due to the completeness property of the eigenstates for each dot.

Each pair of dots is let to have a dipole interaction through the coefficients depending on the type of dots and the energy levels between which the electrons of these dots do a transition. The related Hamiltonian of dipole interaction is given by [17]

$$\mathbb{H}_{r.r} = \sum_{n<m,i<j} \left(\eta_{nij}\hat{\sigma}^{n}_{i,j} + \eta^{*}_{nij}\hat{\sigma}^{n}_{j,i}\right)\left(\eta_{mij}\hat{\sigma}^{m}_{i,j} + \eta^{*}_{mij}\hat{\sigma}^{m}_{j,i}\right). \tag{8}$$

The coefficient $\eta_{nij}$ depends on the strength of the dipole generated by transition of the dot $n$ between two energy levels $i$ and $j$ associated with the atom.

Ignoring interactions between dots and the electromagnetic (EM) field, the basic Hamiltonian is [17]

$$\mathbb{H}_0 = \sum_{n,i} E^{n}_{i}\hat{\sigma}^{n}_{i} + \sum_{\nu} \hbar\Omega_{\nu} a^{\dagger}_{\nu} a_{\nu}, \tag{9}$$

where the zero point energies of radiation fields are dropped. In the latter relation, the value of energy level $i$ of dot $n$ is indicated by $E^{n}_{i}$ and the frequency of mode $\nu$ is indicated by $\Omega_{\nu}$. For each $a$, $b$ and $n$ we assume the convention that if $a<b$ then $E^{n}_{a} < E^{n}_{b}$. The total Hamiltonian can be then represented as $\mathbb{H} = \mathbb{H}_0 + \mathbb{H}_{r.E} + \mathbb{H}_{r.r}$.

## IMAGE HAMILTONIAN

The Image Hamiltonian in the Heisenberg's interaction picture [17] can be expressed as

$$\mathbb{H}_{\text{int}}^{(I)} = \exp\left[\frac{\mathtt{i}}{\hbar}\mathbb{H}_0 t\right]\mathbb{H}_{\text{int}}\exp\left[\frac{-\mathtt{i}}{\hbar}\mathbb{H}_0 t\right], \tag{10}$$

where the total interaction Hamiltonian is

$$\mathbb{H}_{\text{int}} = \mathbb{H}_{r.E} + \mathbb{H}_{r.r}. \tag{11}$$

Our goal is now to calculate (10) term by term. At first, we define $\omega^{n}_{i} \triangleq \frac{E^{n}_{i}}{\hbar}$ and $\xi \triangleq e^{\mathtt{i}t}$.Using (9) we have

$$\exp\left[\frac{\mathtt{i}}{\hbar}\mathbb{H}_0 t\right] = \exp\left[\mathtt{i}\left(\sum_{n,i}\omega^{n}_{i}\hat{\sigma}^{n}_{i} + \sum_{\nu}\Omega_{\nu}a^{\dagger}_{\nu}a_{\nu}\right)t\right] = \prod_{n}\exp\left(\sum_{i}\mathtt{i}\omega^{n}_{i}\hat{\sigma}^{n}_{i}t\right)\exp\left(\sum_{\nu}\mathtt{i}\Omega_{\nu}a^{\dagger}_{\nu}a_{\nu}t\right). \tag{12}$$

In (12), we have used relation $\exp(\hat{A}+\hat{B})=\exp(\hat{A})\exp(\hat{B})$ which is in its turn true if $[\hat{A},\hat{B}]=0$. The definition of the exponential operator and relation (6) result in

$$\exp\left(\sum_i \mathtt{i}\omega_i^n \hat{\sigma}_i^n t\right)=\sum_i \exp\left(\mathtt{i}\omega_i^n t\right)\hat{\sigma}_i^n=\sum_i \xi^{\omega_i^n}\hat{\sigma}_i^n. \quad (13)$$

where $\xi=\exp(\mathtt{i}t)$ is taken for convenience. Also due to the fact that $\left[a_\nu^\dagger a_\nu, a_{\nu'}^\dagger a_{\nu'}\right]=0$, one can conclude that

$$\exp\left(\sum_\nu \mathtt{i}\Omega_\nu a_\nu^\dagger a_\nu t\right)=\prod_\nu \xi^{\Omega_\nu a_\nu^\dagger a_\nu} \quad (14)$$

Combining (12) and (13) and (14) finally leads to

$$\exp\left[\frac{\mathtt{i}}{\hbar}\mathbb{H}_0 t\right]=\prod_n\sum_i \xi^{\omega_i^n}\hat{\sigma}_i^n.\prod_\nu \xi^{\Omega_\nu a_\nu^\dagger a_\nu}. \quad (15)$$

Through a similar procedure, we have

$$\exp\left[-\frac{\mathtt{i}}{\hbar}\mathbb{H}_0 t\right]=\prod_n\sum_i \xi^{-\omega_i^n}\hat{\sigma}_i^n\prod_\nu \xi^{-\Omega_\nu a_\nu^\dagger a_\nu} \quad (16)$$

Using (1),(8),(11) and (15) one can conclude that

$$\exp\left[\frac{\mathtt{i}}{\hbar}\mathbb{H}_0 t\right]\mathbb{H}_{\text{int}}=\prod_s\sum_k \xi^{\omega_k^s}\hat{\sigma}_k^s.\prod_\nu \xi^{\Omega_\nu a_\nu^\dagger a_\nu}\times$$
$$\left[\sum_{n,i<j}\left(\gamma_{nij}\hat{\sigma}_{i,j}^n+\gamma_{nij}^*\hat{\sigma}_{j,i}^n\right)\sum_\nu\left(g_{nij\nu}a_\nu+g_{nij\nu}^*a_\nu^\dagger\right)+\sum_{n<m,i<j,p<q}\left(\eta_{nij}\hat{\sigma}_{i,j}^n+\eta_{nij}^*\hat{\sigma}_{j,i}^n\right)\left(\eta_{mpq}\hat{\sigma}_{p,q}^m+\eta_{mpq}^*\hat{\sigma}_{q,p}^m\right)\right]. \quad (17)$$

To further simplify the calculation of (10), we first notice that for any arbitrary function *f* we have

$$\prod_s\sum_k f(s,k,t)\hat{\sigma}_k^s\hat{\sigma}_{i,j}^n=\prod_{s\neq n}\sum_k f(s,k,t)\hat{\sigma}_k^s\sum_{k'}f(n,k',t)\hat{\sigma}_{k'}^n\hat{\sigma}_{i,j}^n=\prod_{s\neq n}\sum_f f(s,k,t)\hat{\sigma}_f^s f(n,i,t)\hat{\sigma}_{i,j}^n, \quad (18)$$

$$\hat{\sigma}^n_{i,j}\prod_s\sum_k f(s,k,t)\hat{\sigma}^s_k=\hat{\sigma}^n_{i,j}\prod_{s\neq n}\sum_k f(s,k,t)\hat{\sigma}^s_k\sum_{k'}f(n,k',t)\hat{\sigma}^n_{k'}=\prod_{s\neq n}\sum_k f(s,k,t)\hat{\sigma}^s_k\hat{\sigma}^n_{i,j}f(n,j,t). \quad (19)$$

Equation (19) is a result of the fact that $\left[\sigma^s_{a,b},\hat{\sigma}^n_{i,j}\right]=0$ for $s\neq n$, that is the transition operators of the two different dots commute with themselves. Now, let $f(s,k,t)=\xi^{\omega^s_k}$. We then get the following from (17) through (19)

$$\prod_s\sum_k\xi^{\omega^s_k}\hat{\sigma}^s_k\hat{\sigma}^n_{i,j}=\prod_{s\neq n}\sum_k\xi^{\omega^s_k}\hat{\sigma}^s_k\sum_k\xi^{\omega^n_k}\hat{\sigma}^n_k\hat{\sigma}^n_{i,j}=\prod_{s\neq n}\sum_f\xi^{\omega^s_f}\hat{\sigma}^s_f\xi^{\omega^n_i}\hat{\sigma}^n_{i,j}.$$

Using (17), (18) and (6) we arrive at

$$\begin{aligned}&\exp\left[\frac{\mathtt{i}}{\hbar}\mathbb{H}_0 t\right]\mathbb{H}_{\text{int}}=\\&\sum_{n,i<j}\left\{\left[\gamma_{nij}\prod_{s\neq n}\sum_f\xi^{\omega^s_f}\hat{\sigma}^s_f\xi^{\omega^n_i}\hat{\sigma}^n_{i,j}+\gamma^*_{nij}\prod_{s\neq n}\sum_f\xi^{\omega^s_f}\hat{\sigma}^s_f\xi^{\omega^n_j}\hat{\sigma}^n_{j,i}\right]\prod_{\nu'}\xi^{\Omega_{\nu'}a^\dagger_{\nu'}a_{\nu'}}\sum_\nu\left(g_{nij\nu}a_\nu+g^*_{nij\nu}a^\dagger_\nu\right)\right\}+\\&\sum_{n<m,i<j,p<q}\left\{\eta_{nij}\prod_{\nu''}\xi^{\Omega_{\nu''}a^\dagger_{\nu''}a_{\nu''}}\prod_{\substack{s\neq n\\s\neq m}}\sum_f\xi^{\omega^s_f}\hat{\sigma}^s_f\left[\eta_{mpq}\xi^{\omega^n_i+\omega^m_p}\hat{\sigma}^n_{i,j}\hat{\sigma}^m_{p,q}+\ \eta^*_{mpq}\xi^{\omega^n_i+\omega^m_q}\hat{\sigma}^n_{i,j}\hat{\sigma}^m_{q,p}\right]+\right.\\&\left.\eta^*_{nij}\prod_{\nu''}\xi^{\Omega_{\nu''}a^\dagger_{\nu''}a_{\nu''}}\prod_{\substack{s\neq n\\s\neq m}}\sum_f\xi^{\omega^s_f}\hat{\sigma}^s_f\ \left[\eta_{mpq}\xi^{\omega^n_j+\omega^m_p}\hat{\sigma}^n_{j,i}\hat{\sigma}^m_{p,q}+\eta^*_{mpq}\xi^{\omega^n_j+\omega^m_q}\hat{\sigma}^n_{j,i}\hat{\sigma}^m_{q,p}\right]\right\},\end{aligned} \quad (20)$$

Using (7), (16), (19) and (20) we can calculate the image Hamiltonian within the Heisenberg interaction picture as

$$\begin{aligned}&\mathbb{H}_{\text{int}}{}^{(I)}=\exp\left(\frac{\mathtt{i}}{\hbar}\mathbb{H}_0 t\right)\mathbb{H}_{\text{int}}\exp\left(-\frac{\mathtt{i}}{\hbar}\mathbb{H}_0 t\right)=\\&\sum_{n,i<j}\left\{\left(\gamma_{nij}\xi^{\omega^n_{ij}}\hat{\sigma}^n_{i,j}+\gamma^*_{nij}\xi^{\omega^n_{ji}}\hat{\sigma}^n_{j,i}\right)\sum_\nu\left(g_{nij\nu}a_\nu\xi^{-\Omega_\nu}+g^*_{nij\nu}a^\dagger_\nu\xi^{\Omega_\nu}\right)\right\}+\\&\sum_{n<m,i<j,p<q}\left\{\eta_{nij}\eta_{mpq}\xi^{\omega^n_{ij}+\omega^m_{pq}}\hat{\sigma}^n_{i,j}\hat{\sigma}^m_{p,q}+\eta_{nij}\eta^*_{mpq}\xi^{\omega^n_{ij}+\omega^m_{qp}}\hat{\sigma}^n_{i,j}\hat{\sigma}^m_{q,p}+\eta^*_{nij}\eta_{mpq}\xi^{\omega^n_{ji}+\omega^m_{pq}}\hat{\sigma}^n_{j,i}\hat{\sigma}^m_{p,q}+\eta^*_{nij}\eta^*_{mpq}\xi^{\omega^n_{ji}+\omega^m_{qp}}\hat{\sigma}^n_{j,i}\hat{\sigma}^m_{q,p}\right\}\end{aligned} \quad (21)$$

in which $\omega^n_{ij}\triangleq\dfrac{E^n_i-E^n_j}{\hbar}$ represents the transition frequency between $i$-th and $j$-th states of the $n$-th dot.

In derivation of (21), we have used the following relation

$$\prod_{\nu'} \xi^{\Omega_{\nu'} a^{\dagger}_{\nu'} a_{\nu'}} \sum_{\nu} \left( g_{nij\nu} a_{\nu} + g^{*}_{nij\nu} a^{\dagger}_{\nu} \right) \prod_{\nu''} \xi^{-\Omega_{\nu''} a^{\dagger}_{\nu''} a_{\nu''}} = \sum_{\nu} \left( g_{nij\nu} a_{\nu} \xi^{-\Omega_{\nu}} + g^{*}_{nij\nu} a^{\dagger}_{\nu} \xi^{\Omega_{\nu}} \right). \tag{22}$$

Validity of (22) may be established as follows:

$$\begin{aligned}
&\prod_{\nu'} \xi^{\Omega_{\nu'} a^{\dagger}_{\nu'} a_{\nu'}} \sum_{\nu} \left( g_{nij\nu} a_{\nu} + g^{*}_{nij\nu} a^{\dagger}_{\nu} \right) \prod_{\nu''} \xi^{-\Omega_{\nu''} a^{\dagger}_{\nu''} a_{\nu''}} = \\
&\quad \sum_{\nu} \prod_{\nu'} \xi^{\Omega_{\nu'} a^{\dagger}_{\nu'} a_{\nu'}} \left( g_{nij\nu} a_{\nu} + g^{*}_{nij\nu} a^{\dagger}_{\nu} \right) \prod_{\nu''} \xi^{-\Omega_{\nu''} a^{\dagger}_{\nu''} a_{\nu''}} = \\
&\quad \sum_{\nu} \prod_{\nu' \neq \nu} \xi^{\Omega_{\nu'} a^{\dagger}_{\nu'} a_{\nu'}} . \xi^{\Omega_{\nu} a^{\dagger}_{\nu} a_{\nu}} \left( g_{nij\nu} a_{\nu} + g^{*}_{nij\nu} a^{\dagger}_{\nu} \right) \xi^{-\Omega_{\nu} a^{\dagger}_{\nu} a_{\nu}} \prod_{\nu'' \neq \nu} \xi^{-\Omega_{\nu''} a^{\dagger}_{\nu''} a_{\nu''}} = \\
&\quad \sum_{\nu} \prod_{\nu' \neq \nu} \xi^{\Omega_{\nu'} a^{\dagger}_{\nu'} a_{\nu'}} \cdot \prod_{\nu'' \neq \nu} \xi^{-\Omega_{\nu''} a^{\dagger}_{\nu''} a_{\nu''}} \xi^{\Omega_{\nu} a^{\dagger}_{\nu} a_{\nu}} \left( g_{nij\nu} a_{\nu} + g^{*}_{nij\nu} a^{\dagger}_{\nu} \right) \xi^{-\Omega_{\nu} a^{\dagger}_{\nu} a_{\nu}} = \\
&\quad \sum_{\nu} \xi^{\Omega_{\nu} a^{\dagger}_{\nu} a_{\nu}} \left( g_{nij\nu} a_{\nu} + g^{*}_{nij\nu} a^{\dagger}_{\nu} \right) \xi^{-\Omega_{\nu} a^{\dagger}_{\nu} a_{\nu}} = \\
&\quad \sum_{\nu} \left( g_{nij\nu} a_{\nu} \xi^{-\Omega_{\nu}} + g^{*}_{nij\nu} a^{\dagger}_{\nu} \xi^{\Omega_{\nu}} \right).
\end{aligned}$$

**RABI EQUATIONS**

Assuming that $l$ is an integer, and $\mathcal{A} = [r_i]_{1\times k} = [r_1 \; r_2 \; \ldots \; r_k]$ and $\mathcal{F} = [f_i]_{1\times w} = [f_1 \; f_2 \ldots \; f_w]$ are two vectors. We take the following notations for the rest of calculations

$$\begin{aligned}
&\mathcal{A}_{r_n \to i} = \left[ r_1 \; r_2 \; \ldots \; r_{n-1} \; i \; r_{n+1} \; \ldots \; r_k \right], \\
&\mathcal{A}_{r_n \to i, r_m \to j} = \left[ r_1 \; r_2 \; \ldots \; r_{n-1} \; i \; r_{n+1} \; \ldots \; r_{m-1} \; j \; r_{m+1} \; \ldots \; r_k \right], \\
&\left| \mathcal{A} \right\rangle = \bigotimes_{n=1}^{k} \left| \begin{matrix} n \\ r_n \end{matrix} \right\rangle = \left| \begin{matrix} 1 \\ r_1 \end{matrix} \right\rangle \left| \begin{matrix} 2 \\ r_2 \end{matrix} \right\rangle \ldots \left| \begin{matrix} k \\ r_k \end{matrix} \right\rangle, \\
&\left| \mathcal{A}_{r_n \to i} \right\rangle = \bigotimes_{n' \neq n} \left| \begin{matrix} n' \\ r_{n'} \end{matrix} \right\rangle \left| \begin{matrix} n \\ i \end{matrix} \right\rangle, \\
&\left| \mathcal{A}_{r_n \to i, r_m \to j} \right\rangle = \bigotimes_{\substack{n' \neq n \\ n' \neq m}} \left| \begin{matrix} n' \\ r_{n'} \end{matrix} \right\rangle \left| \begin{matrix} n \\ i \end{matrix} \right\rangle \left| \begin{matrix} m \\ j \end{matrix} \right\rangle, \\
&\mathcal{F}_{f_\nu \to f_\nu + l} = \left[ f_1 \; f_2 \; \ldots f_{\nu-1} \; f_\nu + l \; f_{\nu+1} \; \ldots \; f_w \right] \\
&\left| \mathcal{F} \right\rangle = \bigotimes_{\nu=1}^{w} \left| f_\nu \right\rangle = \left| f_1 \right\rangle \left| f_2 \right\rangle \ldots \left| f_w \right\rangle.
\end{aligned}$$

$$\sum_{\mathcal{A},\mathcal{F}}(\cdot)=\sum_{r_1,r_2,\ldots,r_k,f_1,f_2,\ldots f_w}(\cdot),$$
$$\sum_{\mathcal{A}-\{r_n\},\mathcal{F}}(\cdot)=\sum_{r_1,r_2,\ldots r_{n-1},r_{n+1},\ldots r_k,f_1,f_2,\ldots,f_w}(\cdot),$$
$$\sum_{\mathcal{A}-\{r_n,r_m\},\mathcal{F}}(\cdot)=\sum_{r_1,r_2,\ldots r_{n-1},r_{n+1}\ldots r_{m-1},r_{m+1}\ldots r_k,f_1,f_2,\ldots,f_w}(\cdot).$$

The general time-dependent state of the ensemble of dots and photons at any time may be expanded on the basis outer products of individual dot and field eigenstates as

$$\left|\varphi(t)\right\rangle=\sum_{\mathcal{A},\mathcal{F}}\Phi(\mathcal{A},\mathcal{F})\left|\mathcal{A}\right\rangle\left|\mathcal{F}\right\rangle, \tag{23}$$

in which $\left|\mathcal{A}\right\rangle=\overset{k}{\underset{n=1}{\otimes}}\left|\begin{matrix}n\\ r_n\end{matrix}\right\rangle$ denotes the eigenstate of the quantum dot molecule, with $\left|\begin{matrix}n\\ r_n\end{matrix}\right\rangle$ denoting the $r_n$-th state of dot $n$. Also, $\left|\mathcal{F}\right\rangle=\overset{w}{\underset{\nu=1}{\otimes}}\left|f_\nu\right\rangle$ represents the field state when the $\nu$-th mode has photon number $f_\nu$ where $0\le f_\nu\le\infty$. Note that the total number of dots in the dot molecule is denoted by $k$, while the number of energy levels in dot $n$ is given by $B_n$. Also, the total number of modes of EM field is denoted by $w$, so that the indices are bounded by $1\le n\le k, 1\le r_n\le B_n$, and $1\le\nu\le w$.

Equation (23) states that, at any time, $\left|\varphi(t)\right\rangle$ is a superposition of all possible states of the system, including atom and field states, and each state has a time-dependent coefficient equal to $\Phi(\mathcal{A},\mathcal{F})$. According to the time-dependent Schrödinger equation we have

$$\mathtt{i}\hbar\frac{\partial\left|\varphi(t)\right\rangle}{\partial t}=\mathbb{H}_{\text{int}}^{(I)}\left|\varphi(t)\right\rangle, \tag{24}$$

By substituting the representation of (23) in the left side of (24), we obtain

$$\mathtt{i}\hbar\frac{\partial\left|\varphi(t)\right\rangle}{\partial t}=\mathtt{i}\hbar\sum_{\mathcal{A},\mathcal{F}}\dot{\Phi}(\mathcal{A},\mathcal{F})\left|\mathcal{A}\right\rangle\left|\mathcal{F}\right\rangle, \tag{25}$$

Also, we have the following properties for the photon number annihilation and creation operators

$$a_\nu \left| f_\nu \right\rangle = \sqrt{f_\nu} \left| f_\nu - 1 \right\rangle, \tag{26.1}$$

$$a_\nu^\dagger \left| f_\nu \right\rangle = \sqrt{f_\nu + 1} \left| f_\nu + 1 \right\rangle. \tag{26.2}$$

By using (21), (23) and (26), the right side of (24) can be expressed as

$$\begin{aligned}
&\mathbb{H}_{\text{int}}^{(I)} \left| \varphi(t) \right\rangle = \\
&\sum_{n,i<j} \Big\{ \eta_{nij} \xi^{\omega_{ij}^n} [ \sum_{\mathcal{A}-\{r_n\},\mathcal{F}} \sum_{\nu} \Phi(\mathcal{A}_{r_n \to j}, \mathcal{F}_{f_\nu \to f_\nu+1}) g_{nij\nu} \xi^{-\Omega_\nu} \sqrt{f_\nu + 1} \left| \mathcal{F} \right\rangle \left| \mathcal{A}_{r_n \to i} \right\rangle + \\
&\quad \sum_{\mathcal{A}-\{r_n\},\mathcal{F}} \sum_{\nu} \Phi(\mathcal{A}_{r_n \to j}, \mathcal{F}_{f_\nu \to f_\nu-1}) g_{nij\nu}^* \xi^{\Omega_\nu} \sqrt{f_\nu} \left| \mathcal{F} \right\rangle \left| \mathcal{A}_{r_n \to i} \right\rangle ] + \\
&\quad \eta_{nij}^* \xi^{\omega_{ji}^n} [ \sum_{\mathcal{A}-\{r_n\},\mathcal{F}} \sum_{\nu} \Phi(\mathcal{A}_{r_n \to i}, \mathcal{F}_{f_\nu \to f_\nu+1}) g_{nij\nu} \xi^{-\Omega_\nu} \sqrt{f_\nu + 1} \left| \mathcal{F} \right\rangle \left| \mathcal{A}_{r_n \to j} \right\rangle + \\
&\quad \sum_{\mathcal{A}-\{r_n\},\mathcal{F}} \sum_{\nu} \Phi(\mathcal{A}_{r_n \to i}, \mathcal{F}_{f_\nu \to f_\nu-1}) g_{nij\nu}^* \xi^{\Omega_\nu} \sqrt{f_\nu} \left| \mathcal{F} \right\rangle \left| \mathcal{A}_{r_n \to j} \right\rangle ] + \\
&\quad \sum_{n<m,i<j,p<q} \Big\{ \eta_{nij} \eta_{mpq} \xi^{\omega_{ij}^n + \omega_{pq}^m} \sum_{\mathcal{A}-\{r_n,r_m\},\mathcal{F}} \Phi(\mathcal{A}_{r_n \to j, r_m \to q}, \mathcal{F}) \left| \mathcal{F} \right\rangle \left| \mathcal{A}_{r_n \to i, r_m \to p} \right\rangle + \\
&\quad \eta_{nij} \eta_{mpq}^* \xi^{\omega_{ij}^n + \omega_{qp}^m} \sum_{\mathcal{A}-\{r_n,r_m\},\mathcal{F}} \Phi(\mathcal{A}_{r_n \to j, r_m \to p}, \mathcal{F}) \left| \mathcal{F} \right\rangle \left| \mathcal{A}_{r_n \to i, r_m \to q} \right\rangle + \\
&\quad \eta_{nij}^* \eta_{mpq} \xi^{\omega_{ji}^n + \omega_{pq}^m} \sum_{\mathcal{A}-\{r_n,r_m\},\mathcal{F}} \Phi(\mathcal{A}_{r_n \to i, r_m \to q}, \mathcal{F}) \left| \mathcal{F} \right\rangle \left| \mathcal{A}_{r_n \to j, r_m \to p} \right\rangle + \\
&\quad \eta_{nij}^* \eta_{mpq}^* \xi^{\omega_{ji}^n + \omega_{qp}^m} \sum_{\mathcal{A}-\{r_n,r_m\},\mathcal{F}} \Phi(\mathcal{A}_{r_n \to i, r_m \to p}, \mathcal{F}) \left| \mathcal{F} \right\rangle \left| \mathcal{A}_{r_n \to j, r_m \to q} \right\rangle \Big\}.
\end{aligned} \tag{27}$$

Substituting (25) and (27) in (24) leads to

$$\mathrm{i}\hbar \dot{\Phi}(\mathcal{A},\mathcal{F}) = \Sigma_1(\mathcal{A},\mathcal{F}) + \Sigma_2(\mathcal{A},\mathcal{F}) + \Sigma_3(\mathcal{A},\mathcal{F}) + \Sigma_4(\mathcal{A},\mathcal{F}) + \Sigma_5(\mathcal{A},\mathcal{F}) + \Sigma_6(\mathcal{A},\mathcal{F}), \tag{28}$$

in which we have used the following notations for the sake of convenience

$$\begin{aligned}
&\Sigma_1(\mathcal{A},\mathcal{F}) \triangleq \\
&\quad \sum_{n\ ,r_n<j} \gamma_{nr_nj} \xi^{\omega_{r_nj}^n} \left[ \sum_{\nu} g_{nr_nj\nu} \xi^{-\Omega_\nu} \sqrt{f_\nu + 1} \Phi(\mathcal{A}_{r_n \to j}, \mathcal{F}_{f_\nu \to f_\nu+1}) + \sum_{\nu} g_{nr_nj\nu}^* \xi^{\Omega_\nu} \sqrt{f_\nu} \Phi(\mathcal{A}_{r_n \to j}, \mathcal{F}_{f_\nu \to f_\nu-1}) \right],
\end{aligned} \tag{28.1}$$

$$\Sigma_2(\mathcal{A},\mathcal{F}) \triangleq$$

$$\sum_{n\ ,i<r_n} \gamma^*_{nir_n} \xi^{\omega^n_{r_n i}} \left[ \sum_{\nu} g_{nir_n\nu} \xi^{-\Omega_\nu} \sqrt{f_\nu + 1} \Phi(\mathcal{A}_{r_n \to i}, \mathcal{F}_{f_\nu \to f_\nu + 1}) + \sum_{\nu} g^*_{nir_n\nu} \xi^{\Omega_\nu} \sqrt{f_\nu} \Phi(\mathcal{A}_{r_n \to i}, \mathcal{F}_{f_\nu \to f_\nu - 1}) \right], \tag{28.2}$$

$$\Sigma_3(\mathcal{A},\mathcal{F}) \triangleq \sum_{n<m\ ,r_n<j\ ,r_m<q} \eta_{nr_nj} \eta_{mr_mq} \xi^{\omega^n_{r_nj}+\omega^m_{r_mq}} \Phi(\mathcal{A}_{r_n\to j, r_m\to q}, \mathcal{F}), \tag{28.3}$$

$$\Sigma_4(\mathcal{A},\mathcal{F}) \triangleq \sum_{n<m\ ,r_n<j\ ,p<r_m} \eta_{nr_nj} \eta^*_{mpr_m} \xi^{\omega^n_{r_nj}+\omega^m_{r_mp}} \Phi(\mathcal{A}_{r_n\to j, r_m\to p}, \mathcal{F}), \tag{28.4}$$

$$\Sigma_5(\mathcal{A},\mathcal{F}) \triangleq \sum_{n<m\ ,i<r_n\ ,r_m<q} \eta^*_{nir_n} \eta_{mr_mq} \xi^{\omega^n_{r_ni}+\omega^m_{r_mq}} \Phi(\mathcal{A}_{r_n\to i, r_m\to q}, \mathcal{F}), \tag{28.5}$$

$$\Sigma_6(\mathcal{A},\mathcal{F}) \triangleq \sum_{n<m\ ,i<r_n\ ,p<r_m} \eta^*_{nir_n} \eta^*_{mpr_m} \xi^{\omega^n_{r_ni}+\omega^m_{r_mp}} \Phi(\mathcal{A}_{r_n\to i, r_m\to p}, \mathcal{F}), \tag{28.6}$$

Within RWA, $\Sigma_3(\mathcal{A},\mathcal{F})$ and $\Sigma_6(\mathcal{A},\mathcal{F})$ may be ignored.

## ANALYTICAL SOLUTION FOR THE SINGLE MODE FIELD

In this section, we assume that we have a single-mode EM field with photon number *f* and the RWA is valid. These assumptions lead to the following relations

$$i\hbar\dot{\Phi}(\mathcal{A},f) = \Sigma_1(\mathcal{A},f) + \Sigma_2(\mathcal{A},f) + \Sigma_4(\mathcal{A},f) + \Sigma_5(\mathcal{A},f), \tag{29.1}$$

$$\Sigma_1(\mathcal{A},f) = \sum_{n\ ,r_n<j} \gamma_{nr_nj} g_{nr_nj} \xi^{-\Delta^n_{r_nj}} \sqrt{f+1} \Phi(\mathcal{A}_{r_n\to j}, f+1)\ , \tag{29.2}$$

$$\Sigma_2(\mathcal{A},f) = \sum_{n\ ,i<r_n} \gamma^*_{nir_n} g^*_{nir_n} \xi^{\Delta^n_{ir_n}} \sqrt{f} \Phi(\mathcal{A}_{r_n\to i}, f-1), \tag{29.3}$$

$$\Sigma_3(\mathcal{A},f) = 0, \tag{29.4}$$

$$\Sigma_4(\mathcal{A},f) = \sum_{n<m\ ,r_n<j\ ,p<r_m} \eta_{nr_nj} \eta^*_{mpr_m} \xi^{\Delta^m_{pr_m}-\Delta^n_{r_nj}} \Phi(\mathcal{A}_{r_n\to j, r_m\to p}, f), \tag{29.5}$$

$$\Sigma_5(\mathcal{A},f) = \sum_{n<m\ ,i<r_n\ ,r_m<q} \eta^*_{nir_n} \eta_{mr_mq} \xi^{\Delta^m_{r_mq}-\Delta^n_{ir_n}} \Phi(\mathcal{A}_{r_n\to i, r_m\to q}, f), \tag{29.6}$$

$$\Sigma_6(\mathcal{A},\mathcal{F}) = 0, \tag{29.7}$$

in which $\Delta_{ab}^{n} \triangleq \Omega - \omega_{ab}^{n}$. Clearly, the above relations provides a distinct differential equation for each $\mathcal{A}$. Since the whole number of all possible vectors $\mathcal{A} = [r_1\ r_2\ \ldots\ r_k]$ such that $1 \le r_n \le \mathrm{B}_n$ is $N = \prod_{i=1}^{k} B_i$, equation (29) leads to a set of $N$ equations. As an example, consider a single-mode EM field $w = 1$, and two dots $k = 2$ each having two energy levels. Thus, $B_1 = 2, B_2 = 2$ and $N = \mathrm{B}_1\mathrm{B}_2 = 4$. If we assume that $\mathcal{A}_1 = [1\ 1], \mathcal{A}_2 = [1\ 2], \mathcal{A}_3 = [2\ 1], \mathcal{A}_4 = [2\ 2]$, the total set of equations generated by (29) take the form

$$i\hbar\dot{\Phi}(\mathcal{A}_1, f) = \wp_1 \xi^{-\Delta_2}\sqrt{f+1}\Phi(\mathcal{A}_2, f+1) + \wp_1 \xi^{-\Delta_1}\sqrt{f+1}\Phi(\mathcal{A}_3, f+1), \tag{30.1}$$

$$i\hbar\dot{\Phi}(\mathcal{A}_2, f) = \wp_1 \xi^{\Delta_2}\sqrt{f}\Phi(\mathcal{A}_1, f-1) + \wp_1 \xi^{-\Delta_1}\sqrt{f+1}\Phi(\mathcal{A}_4, f+1) + \wp_2 \xi^{\Upsilon}\Phi(\mathcal{A}_3, f), \tag{30.2}$$

$$i\hbar\dot{\Phi}(\mathcal{A}_3, f) = \wp_1 \xi^{\Delta_1}\sqrt{f}\Phi(\mathcal{A}_1, f-1) + \wp_1 \xi^{-\Delta_2}\sqrt{f+1}\Phi(\mathcal{A}_4, f+1) + \wp_2 \xi^{-\Upsilon}\Phi(\mathcal{A}_2, f), \tag{30.3}$$

$$i\hbar\dot{\Phi}(\mathcal{A}_4, f) = \wp_1 \xi^{\Delta_1}\sqrt{f}\Phi(\mathcal{A}_2, f-1) + \wp_1 \xi^{\Delta_2}\sqrt{f}\Phi(\mathcal{A}_3, f-1), \tag{30.4}$$

in which $\wp_1 = \gamma_{abc} g_{def} = \gamma_{abc}^{*} g_{def}^{*}, \wp_2 = \eta_{abc}\eta_{def}^{*} = \eta_{abc}^{*}\eta_{def}, \Delta_1 \triangleq \Omega_1 - \omega_{12}^{1}$, $\Delta_2 \triangleq \Omega_1 - \omega_{12}^{2}$ and $\Upsilon \triangleq \omega_{12}^{1} - \omega_{12}^{2} = \Delta_2 - \Delta_1$. One may rewrite (30) as

$$\begin{aligned}
i\hbar\dot{\Phi}(\mathcal{A}_1, f) &= \wp_1 \xi^{-\Delta_2}\sqrt{f+1}\Phi(\mathcal{A}_2, f+1) + \wp_1 \xi^{-\Delta_1}\sqrt{f+1}\Phi(\mathcal{A}_3, f+1), \\
i\hbar\dot{\Phi}(\mathcal{A}_2, f+1) &= \wp_1 \xi^{\Delta_2}\sqrt{f+1}\Phi(\mathcal{A}_1, f) + \wp_1 \xi^{-\Delta_1}\sqrt{f+2}\Phi(\mathcal{A}_4, f+2) + \wp_2 \xi^{\Upsilon}\Phi(\mathcal{A}_3, f+1), \\
i\hbar\dot{\Phi}(\mathcal{A}_3, f+1) &= \wp_1 \xi^{\Delta_1}\sqrt{f+1}\Phi(\mathcal{A}_1, f) + \wp_1 \xi^{-\Delta_2}\sqrt{f+2}\Phi(\mathcal{A}_4, f+2) + \wp_2 \xi^{-\Upsilon}\Phi(\mathcal{A}_2, f+1), \\
i\hbar\dot{\Phi}(\mathcal{A}_4, f+2) &= \wp_1 \xi^{\Delta_1}\sqrt{f+2}\Phi(\mathcal{A}_2, f+1) + \wp_1 \xi^{\Delta_2}\sqrt{f+2}\Phi(\mathcal{A}_3, f+1).
\end{aligned}$$

Using the notation $\Theta_{1,f} \triangleq \Phi(\mathcal{A}_1, f), \Theta_{2,f} \triangleq \Phi(\mathcal{A}_2, f+1), \Theta_{3,f} \triangleq \Phi(\mathcal{A}_3, f+1)$ and $\Theta_{4,f} \triangleq \Phi(\mathcal{A}_4, f+2)$ and multiplying the resultant equations by $1, \xi^{-\Delta_{12}^{2}}, \xi^{-\Delta_{12}^{1}}$ and $\xi^{-\Delta_{12}^{1}-\Delta_{12}^{2}}$, respectively, results in

$$\begin{aligned}
i\hbar\dot{\Theta}_{1,f} &= \wp_1\sqrt{f+1}\xi^{-\Delta_{12}^{2}}\Theta_{2,f} + \wp_1\sqrt{f+1}\xi^{-\Delta_{12}^{1}}\Theta_{3,f}, \\
i\hbar\xi^{-\Delta_{12}^{2}}\dot{\Theta}_{2,f} &= \wp_1\sqrt{f+1}\Theta_{1,f} + \wp_1\sqrt{f+2}\xi^{-\Delta_{12}^{1}-\Delta_{12}^{2}}\Theta_{4,f} + \wp_2\xi^{-\Delta_{12}^{1}}\Theta_{3,f}, \\
i\hbar\xi^{-\Delta_{12}^{1}}\dot{\Theta}_{3,f} &= \wp_1\sqrt{f+1}\Theta_{1,f} + \wp_1\sqrt{f+2}\xi^{-\Delta_{12}^{2}-\Delta_{12}^{1}}\Theta_{4,f} + \wp_2\xi^{-\Delta_{12}^{2}}\Theta_{2,f}, \\
i\hbar\xi^{-\Delta_{12}^{2}-\Delta_{12}^{1}}\dot{\Theta}_{4,f} &= \wp_1\sqrt{f+2}\xi^{-\Delta_{12}^{2}}\Theta_{2,f} + \wp_1\sqrt{f+2}\xi^{-\Delta_{12}^{1}}\Theta_{3,f}.
\end{aligned}$$

Now, we define $A \triangleq \Theta_{1,f}, B \triangleq \xi^{-\Delta_{12}^2}\Theta_{2,f}, C \triangleq \xi^{-\Delta_{12}^1}\Theta_{3,f}, D \triangleq \xi^{-\Delta_{12}^2-\Delta_{12}^1}\Theta_{4,f}$, and use the property that if $X(t) \triangleq \xi^{\Delta}\Theta(t)$ then $\xi^{\Delta}\Theta' = X' - i\Delta X$, and rewrite the above set of equations into the following matrix form

$$\mathrm{i}\hbar\Psi' = \Xi\Psi, \tag{30.5}$$

$$\Psi \triangleq \begin{bmatrix} A & B & C & D \end{bmatrix}^T,$$

$$\Xi \triangleq \begin{bmatrix} 0 & \wp_1\sqrt{f+1} & \wp_1\sqrt{f+1} & 0 \\ \wp_1\sqrt{f+1} & \hbar\Delta_{12}^2 & \wp_2 & \wp_1\sqrt{f+2} \\ \wp_1\sqrt{f+1} & \wp_2 & \hbar\Delta_{12}^1 & \wp_1\sqrt{f+2} \\ 0 & \wp_1\sqrt{f+2} & \wp_1\sqrt{f+2} & \hbar\left(\Delta_{12}^1+\Delta_{12}^2\right) \end{bmatrix}.$$

Equation (30.5) consists of a set of linear differential equations that can be solved using standard Laplace Transform techniques either by hand, or more conveniently using symbolic computation softwares such as Mathematica or Maple.

Now, we present a proof that an analytical solution can be always found for (29), that is for single mode quantized radiation fields, an analytical solution always exists regardless of the number of dots. At first we need to present a definition for the precedence:

Assume that $\mathcal{A} = \mathcal{A}_x = \begin{bmatrix} r_1 & r_2 & \dots & r_k \end{bmatrix}$ and $\mathcal{A}' = \mathcal{A}_y = \begin{bmatrix} s_1 & s_2 & \dots & s_k \end{bmatrix}$; then if for one $n$ and for any $i \neq n$, relations $r_n < s_n$ and $r_i = s_i$ are satisfied, we say that $\mathcal{A}$ precedes $\mathcal{A}'$, here denoted by either $\mathcal{A} \prec \mathcal{A}'$ or $\mathcal{A}' \succ \mathcal{A}$. We furthermore define $\Lambda_{x,y} \triangleq -\Delta_{r_n,s_n}^n$ in which $\Delta_{ab}^n \triangleq \Omega - \omega_{ab}^n$. For example, if $\mathcal{A} = \mathcal{A}_1 = \begin{bmatrix} 1 & 3 & 4 & 5 & 2 \end{bmatrix}$, and $\mathcal{A}' = \mathcal{A}_2 = \begin{bmatrix} 1 & 3 & 4 & 2 & 2 \end{bmatrix}$, then $\mathcal{A}' \prec \mathcal{A}$ and therefore $\Lambda_{2,1} \triangleq -\Delta_{2,5}^4$.

We can now rewrite (29.1) as

$$\mathrm{i}\hbar\dot{\Phi}(\mathcal{A}_i, f) = \sum_{j=1}^{N} M_{i,j,f}\,\xi^{\chi_{i,j}}\Phi(\mathcal{A}_j, f + z_{i,j}) \tag{31}$$

in which $M_{i,j,f}$ are constants pertaining to interaction coefficients and the field photon number . The parameter $z_{i,j}$ take on either of the values 0, −1 or +1 depending on the relation between $\mathcal{A}_i$ and $\mathcal{A}_j$. Specifically, as observed in (29.1) and (29.2)

$$z_{i,j} = -1 \text{ ,if } \mathcal{A}_i \succ \mathcal{A}_j, \tag{32.1}$$

$$z_{i,j} = +1 \text{, if } \mathcal{A}_i \prec \mathcal{A}_j. \tag{32.2}$$

Furthermore, in the two latter cases, $\chi_{i,j}$ can be obtained as

$$\chi_{i,j} = \Lambda_{i,j} \text{, if } \mathcal{A}_i \prec \mathcal{A}_j, \tag{32.3}$$

$$\chi_{i,j} = \Lambda_{j,i} \text{, if } \mathcal{A}_i \succ \mathcal{A}_j. \tag{32.4}$$

For instance, in the case of the previous example, using equations (30.1) to (30.4) we have

$$M_{2,1,f} = \wp_1\sqrt{f}, M_{2,2,f} = 0, M_{2,3,f} = \wp_2, M_{2,4,f} = \wp_1\sqrt{f+1},$$
$$\chi_{2,1} = \Delta_2, \chi_{2,2} = 0, \chi_{2,3} = \Upsilon, \chi_{2,4} = \text{-}\Delta_1,$$
$$\mathcal{A}_1 \prec \mathcal{A}_2, z_{2,1} = -1, \Lambda_{1,2} = -\Delta_{1,2}^2.$$

Now we present the following two lemmas, which allow the conversion of (29) into an analytically-solvable form.

*Lemma 1.* Equation (31) is equivalent to

$$\mathrm{i}\hbar\dot{\Phi}(\mathcal{A}_i, f+\varsigma_i) = \sum_{k=1}^{N} M_{i,k,f+\varsigma_k}\xi^{\chi_{i,k}}\Phi(\mathcal{A}_k, f+\varsigma_k), \tag{33}$$

in which we have $\varsigma_1 = 0$, and $\varsigma_i = \varsigma_j + 1$, if $\mathcal{A}_i \succ \mathcal{A}_j$, $1 \le i, j \le N$.

*Proof.* We take an induction approach to prove Lemma 1. Assume that the Lemma is true for $1 \le i, j \le p$, that is

$$\mathrm{i}\hbar\dot{\Phi}(\mathcal{A}_i, f+\varsigma_i) = \sum_{k=1}^{N} M_{i,k,f+\varsigma_i}\xi^{\chi_{i,k}}\Phi(\mathcal{A}_k, f+\varsigma_k), \tag{34.1}$$

in which $\varsigma_i = \varsigma_j + 1$, if $\mathcal{A}_i \succ \mathcal{A}_j$, and $1 \le i, j \le p$. Now, we must prove that the Lemma also holds for $1 \le i, j \le p+1$. This requires us to prove (34.1) for the two following cases

(i) $\varsigma_i = \varsigma_{p+1} + 1$, if $\mathcal{A}_i \succ \mathcal{A}_{p+1}$, $(1 \le i \le p)$ and
(ii) $\varsigma_{p+1} = \varsigma_j + 1$, if $\mathcal{A}_{p+1} \succ \mathcal{A}_j$, $(1 \le j \le p+1)$.

Here, we just prove (i); the proof of (ii) is similar. If $\mathcal{A}_i \succ \mathcal{A}_{p+1}$, due to (32) we have

$$z_{i,p+1} = -1 \tag{34.2}$$

Replacing $f$ with $f+\varsigma_i$ in (31) leads to

$$\mathrm{i}\hbar\dot{\Phi}(\mathcal{A}_i, f+\varsigma_i) = \sum_{k=1}^{N} M_{i,k,f+\varsigma_i}\xi^{\chi_{i,k}}\Phi(\mathcal{A}_k, f+z_{i,k}+\varsigma_i). \tag{34.3}$$

A comparison between (34.1) and (34.3) shows that for the summation index $k = p+1$ we have

$$\varsigma_{p+1} = z_{i,p+1} + \varsigma_i. \tag{34.4}$$

Using (34.2) and (34.4) leads to (i). □

For instance, In the case of the previous example we had $\mathcal{A}_1 = [1\ 1], \mathcal{A}_2 = [1\ 2], \mathcal{A}_3 = [2\ 1]$ and $\mathcal{A}_4 = [2\ 2]$. So, $\varsigma_1 = 0$, $\varsigma_2 = \varsigma_1 + 1 = 1$, $\varsigma_3 = \varsigma_1 + 1 = 1$ and $\varsigma_4 = \varsigma_2 + 1 = 2$. Equations (30.1-4) describe the system in agreement with Lemma 1.

If we now define $\Theta_{i,f} \triangleq \Phi(\mathcal{A}_i, f+\varsigma_i)$, equation (33) can be expressed as

$$\mathrm{i}\hbar\dot{\Theta}_{i\,f} = \sum_{j=1}^{N} \wp_{ijf}\, \xi^{\chi_{i,j}}\, \Theta_{j\,f}\,, \tag{35}$$

where $\wp_{ijf} \triangleq M_{i,j,f+\varsigma_i}$.

*Lemma* 2. Equation (35) is equivalent to

$$\mathrm{i}\hbar\Gamma_i\dot{\Theta}_{i,f} = \sum_j \wp_{ijf}\Gamma_j\Theta_{j,f}. \tag{36}$$

in which we have $\Gamma_1 = 1$ and $\Gamma_j = \Gamma_i\xi^{\Lambda_{i,j}}$ if $\mathcal{A}_i \prec \mathcal{A}_j,\ 1 \le i, j \le N.$

*Proof.* Again we use induction to prove Lemma 2. Assume that the assumption holds for $1 \le i, j \le p$, that is

$$\mathrm{i}\hbar\Gamma_i\dot{\Theta}_{i,f} = \sum_k \wp_{ikf}\Gamma_k\Theta_{k,f}, \tag{37.1}$$

in which $\Gamma_j = \Gamma_i\xi^{\Lambda_{i,j}}$, if $\mathcal{A}_i \prec \mathcal{A}_j,\ 1 \le i, j \le p.$ Now, we must prove that it is also true for $1 \le i, j \le p+1$. That is, we must prove (36) for the two following cases

(iii) $\Gamma_{p+1} = \Gamma_i\xi^{\Lambda_{i,p+1}}$, if $\mathcal{A}_{p+1} \succ \mathcal{A}_i,\ 1 \le i \le p,$
(iv) $\Gamma_j = \Gamma_{p+1}\xi^{\Lambda_{p+1,j}}$, if $\mathcal{A}_j \succ \mathcal{A}_{p+1},\ 1 \le j \le p+1,$

Here, we only prove (iii); the proof of (iv) is similar.

Assume that $\mathcal{A}_{p+1} \succ \mathcal{A}_i$ .Using (32.3) we have

$$\chi_{i,p+1} = \Lambda_{i,p+1}, \tag{37.2}$$

If we multiply both sides of (35) by $\Gamma_i$ , we obtain

$$\mathrm{i}\hbar\Gamma_i\dot{\Theta}_{i,f} = \sum_{k=1}^{N}\wp_{ikf}\Gamma_i\xi^{\chi_{i,k}}\Theta_{k,f}, \tag{37.3}$$

A comparison between (37.1) and (37.3) shows that for the summation index $k = p+1$ we have

$$\Gamma_{p+1} = \xi^{\chi_{i,p+1}}\Gamma_i, \tag{37.4}$$

Through (37.2) and (37.4), (iii) is proved. □

For instance, In the case of the previous example we had $\mathcal{A}_1 = [1\ 1], \mathcal{A}_2 = [1\ 2], \mathcal{A}_3 = [2\ 1]$ and $\mathcal{A}_4 = [2\ 2]$. So, $\Gamma_1 = 1$, $\Lambda_{1,2} = -\Delta_{1,2}^2$ , $\Gamma_2 = \Gamma_1\xi^{\Lambda_{1,2}} = \xi^{-\Delta_{1,2}^2}$, $\Lambda_{1,3} = -\Delta_{1,2}^1$ , $\Gamma_3 = \Gamma_1\xi^{\Lambda_{1,3}} = \xi^{-\Delta_{1,2}^1}$, $\Lambda_{2,4} = -\Delta_{1,2}^1$ and $\Gamma_4 = \Gamma_2\xi^{\Lambda_{2,4}} = \xi^{-\Delta_{1,2}^1-\Delta_{1,2}^2}$ . Hence, through Lemma 2 it is seen that in (36) there is a number such as $\theta_i$ for which we have $\Lambda_{2,4} = -\Delta_{1,2}^1$ . If we define $\Gamma_i = \xi^{\theta_i}$, (36) can be written as

$$\mathrm{i}\hbar X'_{i,f} = \sum_j \wp_{ijf}X_{j,f} - \hbar\theta_i X_{i,f},$$

which is equivalent to the following matrix form

$$\mathrm{i}\hbar\Psi' = \Xi\Psi, \tag{38}$$

where $\Psi \triangleq \left[\Psi_i\right]_{N\times 1}, \Psi_i = X_{i,f}, \Xi \triangleq \left[\Xi_{i,j}\right]_{N\times N}$, and $\Xi_{i,j} = \wp_{ijf} - \hbar\theta_i\delta_{i,j}$.

Equation (38) consists of a set of linear differential equations that can be solved using the Laplace Transform.

## NUMERICAL EXAMPLE

Consider two two-level dots with the energy levels *a* and *b* for the first dot, and *c* and *d* for the second one as shown in Fig. 1, in a cavity with two modes of radiation field. We can write the time-dependent Schrödinger's equation for the system as

$$i\hbar\frac{\partial|\varphi(t)\rangle}{\partial t}=H_{\text{int}}^{(I)}|\varphi(t)\rangle,$$

$$|\varphi(t)\rangle=\sum\left[\varphi_{a,c,n_1,n_2}|a,c,n_1,n_2\rangle+\varphi_{a,d,n_1,n_2}|a,d,n_1,n_2\rangle+\varphi_{b,c,n_1,n_2}|b,c,n_1,n_2\rangle+\varphi_{b,d,n_1,n_2}|b,d,n_1,n_2\rangle\right].$$

where $n_1$ and $n_2$ correspond to the photon numbers of the two dots. Assume furthermore that $\hbar\Delta_{1m}=\hbar\Omega_m-(E_a-E_b),\hbar\Delta_{2m}=\hbar\Omega_m-(E_c-E_d),\hbar q=(E_a-E_b)-(E_c-E_d),$ where $\Omega_i$ is the $i$-th mode frequency, $E_k$ is the energy of the level $k$ ($k=a,b,c,d$) and $\hbar$ is the Plank's constant. The following system of equations can be obtained owing to RWA by (28)

$$\begin{aligned}i\dot{A}(n_1,n_2)&=g_{11}e^{-it\Delta_{11}}C(n_1+1,n_2)\sqrt{n_1+1}+g_{21}e^{-it\Delta_{21}}B(n_1+1,n_2)\sqrt{n_1+1}\\&+g_{12}e^{-it\Delta_{12}}C(n_1,n_2+1)\sqrt{n_2+1}+g_{22}e^{-it\Delta_{22}}B(n_1,n_2+1)\sqrt{n_2+1},\end{aligned}\tag{39.1}$$

$$\begin{aligned}i\dot{B}(n_1,n_2)&=g_{11}e^{-it\Delta_{11}}D(n_1+1,n_2)\sqrt{n_1+1}+g_{21}e^{it\Delta_{21}}A(n_1-1,n_2)\sqrt{n_1}\\&+g_{12}e^{-it\Delta_{12}}D(n_1,n_2+1)\sqrt{n_2+1}+g_{22}e^{it\Delta_{22}}A(n_1,n_2-1)\sqrt{n_2}+\wp e^{itq}C(n_1,n_2),\end{aligned}\tag{39.2}$$

$$\begin{aligned}i\dot{C}(n_1,n_2)&=g_{21}e^{-it\Delta_{21}}D(n_1+1,n_2)\sqrt{n_1+1}+g_{11}e^{it\Delta_{11}}B(n_1-1,n_2)\sqrt{n_1}\\&+g_{22}e^{-it\Delta_{22}}D(n_1,n_2+1)\sqrt{n_2+1}+g_{12}e^{it\Delta_{12}}B(n_1,n_2-1)\sqrt{n_2+1}+\wp e^{-itq}B(n_1,n_2),\end{aligned}\tag{39.3}$$

$$\begin{aligned}i\dot{D}(n_1,n_2)&=g_{21}e^{it\Delta_{21}}C(n_1-1,n_2)\sqrt{n_1}+g_{11}e^{it\Delta_{11}}B(n_1-1,n_2)\sqrt{n_1}\\&+g_{22}e^{it\Delta_{22}}C(n_1,n_2-1)\sqrt{n_2}+g_{12}e^{it\Delta_{12}}B(n_1,n_2-1)\sqrt{n_2}.\end{aligned}\tag{39.4}$$

where $g_{mn}$ is the coupling coefficient associated with the $m$-th dot and the $n$-the mode of the field in the cavity, $\wp$ is the dipole-dipole coupling constant. Notice that we have used a slightly different notation for simplicity, given by $A(n_1,n_2)=\varphi_{a,c,n_1,n_2},B(n_1,n_2)=\varphi_{a,d,n_1,n_2},C(n_1,n_2)=\varphi_{b,c,n_1,n_2},D(n_1,n_2)=\varphi_{b,d,n_1,n_2}$.

If the above system contained one cavity mode, we could use analytical methods such as the Laplace transform to solve it as discussed before. In the case of two or more cavity modes, the system is more complex and finding an analytical solution seems unlikely. Hence we turn to the numerical solution.

We can solve the above system with a simple finite difference scheme. Forward difference scheme is used to approximate the first order derivatives $\frac{df(t)}{dt}\equiv g(t)\approx\frac{f^{n+1}-f^n}{\Delta t}$, where $n$ is the index of the time step.

Therefore the discretization is as $f^{n+1} = g^n \Delta t + f^n$. As $n_1$ and $n_2$ increase, the algorithm becomes more and more time-consuming. If $D(n_1,n_2)$ is desired in (39), $D(i,j)$ will be required if $i$ and $j$ satisfy $i+j=n_1+n_2+1$. Also $B(i,j)$ and $C(i,j)$ with $i+j=n_1+n_2$ and $A(i,j)$ with $i+j=n_1+n-1$ would be needed. The total number of equations to be solved is therefore $4(n_1+n_2)$. Table 1 and table 2 show the required variables for $n_1+n_2=2$ and $n_1+n_2=3$.

We solve the system of equations for the simplest case, where $n_1+n_2=2$ as in Table 1. For the first example, the we take $\wp = g_{i1} = \Omega_1 / 2\pi = \omega_{ab} / 2 = \omega_{cd} / 2 = 2 \times 10^{13}$ Hz and $g_{i2} = \Omega_2 / 2\pi = 3 \times 10^{13}$ Hz ($i$=1,2). These parameters represent a system that consists of two similar two-level dots with the same value of transition energy. The coupling coefficients of each dot to different cavity modes will be however taken to be different, so that $g_{ij}$ and $g_{ik}$ are not the same. The associated simulation time step is $\Delta t = 5 \times 10^{-19}$ sec, and the initial conditions were $A(0,0) = 1$, $D(1,1) = D(2,0) = D(0,2) = C(1,0) = C(0,1) = B(1,0) = B(0,1) = 0$. The wave form for the absolute value of $|A(0,0)|$ is plotted in Fig. 2.

The convergence of the finite difference scheme is an important issue, which needs particular attention. The test for the validity of the method can be the energy conservation law. As the system is lossless, the overall energy in the system must be a fixed value; hence, if the time step is not chosen small enough, the obtained wave forms will gradually decay. In Fig. 3, we show the $|A(0,0)|$ for three different $\Delta t$, where only the last one with $\Delta t=5\times10^{-18}$ renders acceptable.

In the previous example, we considered two similar dots and two different cavity modes. Had we considered two different dots with $g_{ij}\neq g_{kj}$, the results would be slightly altered, however. The common point is that the conservation of energy must be satisfied through the temporal evolution of the states. $\Delta_{mn}$ in (39) appear in the exponents, so even a slight change in their values can lead to frequency shifts in the Fourier transform of the results. On the other hand, $g_{ij}$ exhibit less effect on the frequency shifts and mainly contribute to the amplitudes of the frequency components.

**Concurrence and degree of entanglement**

To measure the degree of entanglement quantitatively, several methods might be used; however, we use the concurrence as mentioned in [21]. According to [21], concurrency for the previously mentioned system of two two-level dots in a cavity with two modes of EM field can be expressed as

$$\lambda = \sqrt{\sum_{n_2,m_2=1}^{N_2} \sum_{n_1,m_1=1}^{N_1} \left|A(n_1,n_2)D(m_1,m_2) - B(n_1,n_2)C(m_1,m_2) - B(m_1,m_2)C(n_1,n_2) + A(m_1,m_2)D(n_1,n_2)\right|^2}.$$

We have considered that $\wp = g_{11} = g_{21} = \Omega_1 = \omega_{ab} = 0.2\times 10^{14}$ and $g_{12} = g_{22} = \Omega_2 = \omega_{cd} = 0.3\times 10^{14}$ where $\omega_{ab} = \left(E_a - E_b\right)/\hbar$ and $\omega_{cd} = \left(E_c - E_d\right)/\hbar$. The total number of equations to be solved in (39) to obtain the time evolution of $D(n_1,n_2)$ would be equal to $4(n_1+n_2)$. Hence, each $N_{sum} = n_1+n_2$ is associated with a set of equations such as (39). Note that we should have $N_{sum} \geq 2$ for the existence of both of the EM field modes. To determine $N_1$ and $N_2$ in the definition of concurrency, one can change $N_{sum}$ such that $2 \leq N_{sum} \leq N_{max}$. For each $N_{sum}$, we have assumed all variables equal to zero at the start of the simulation except $A(1, N_{sum}-1)$, which was set to unity. Fig. 4 shows the time evolution of concurrency $\lambda$ evaluated for $N_{max} = 3$. As expected, the system rapidly undergoes entanglement in the first 20fs, and the degree of entanglement remains within roughly 10% of the average value.

**CONCLUSIONS**

We followed the JCPM model to obtain a set of linearly coupled equations describing the interaction of an electromagnetic field of arbitrary number of modes with a collection of dots each having arbitrary energy levels, acting as a quantum dot molecule. We simplified the equations for a single mode field and applied the RWA. Then, we investigated the obtained equations and presented a method to solve them analytically after analysing a particular example. The increased number of cavity modes causes the solution to become more complex and analytical methods such as the Laplace transform method, cannot be found as easily as before. The system governing equations are achieved and it is found that the sum of the photon number of the EM modes determines the total number of equations to be solved. The finite difference method is used as the easiest way to solve the system. However, the latter method is time consuming and its accuracy is influenced by the time step size. Finally we demonstrated the entanglement of the system of two two-level dots and two modes of the electromagnetic fields in a lossless cavity, and used concurrency as the measure to evaluate multi-particle entanglement.

**FIGURE AND TABLE CAPTIONS**

Table 1. The desired and required variables in (39) for $n_1+n_2=2$.

Table 2. The desired and required variables in (39) for $n_1+n_2=3$.

Figure 1. The investigated system. Two two-level dots are considered with different energy levels and consequently, different energy gaps. Two modes of the electromagnetic field are also accounted for to be entangled with the two dots.

Figure 2. Time evolution of the absolute value of $|A(0,0)|$.

Figure 3. Up: Effect of the size of time step on the accuracy of the result of finite difference method. Up: $\Delta t=5\times10^{-19}$. Middle: $\Delta t=5\times10^{-18}$. Bottom: $\Delta t=5\times10^{-17}$. In all of the above figures, the horizontal axis is the time axis, the vertical axis shows the values of $|A(0,0)|$ and the simulation is run for the overall time of $T=10^{-12}$s. Due to the energy conservation law, the wave form average should neither increase, nor decrease as the time goes on. The decay in larger time steps is a consequence of the innate error in the finite difference method.

Figure 4. Time evolution of concurrency of the system of two two-level quantum dots in a cavity with two EM modes.

Table 1.

| **Desired** | **Required** |
|---|---|
| $D(1,1)$ | $C(1,0),\ C(0,1),\ B(1,0),\ B(0,1)$ |
| $C(1,0)$ | $A(0,0),\ D(2,0),\ D(1,1),\ B(1,0)$ |
| $C(0,1)$ | $A(0,0),\ D(0,2),\ D(1,1),\ B(0,1)$ |
| $B(1,0)$ | $A(0,0),\ D(2,0),\ D(1,1),\ C(1,0)$ |
| $B(0,1)$ | $A(0,0),\ D(0,2),\ D(1,1),\ C(0,1)$ |
| $A(0,0)$ | $C(1,0),\ C(0,1),\ B(1,0), C(0,1)$ |
| $D(2,0)$ | $B(1,0),\ C(1,0)$ |
| $D(0,2)$ | $B(0,1),\ C(0,1)$ |

Table 2

| **Desired** | **Required** |
|---|---|
| $D(2,1)$ | $C(2,0),\ C(1,1),\ B(2,0),\ B(1,1)$ |
| $C(2,0)$ | $A(1,0),\ D(3,0),\ D(2,1),\ B(2,0)$ |
| $C(1,1)$ | $A(1,0),\ A(0,1),\ D(1,2),\ D(2,1),\ B(1,1)$ |
| $B(2,0)$ | $A(1,0),\ D(3,0),\ D(2,1),\ C(2,0)$ |
| $B(1,1)$ | $A(1,0),\ A(0,1),\ D(1,2),\ D(2,1),\ C(1,1)$ |
| $A(1,0)$ | $C(2,0),\ C(1,1),\ B(2,0),\ B(1,1)$ |
| $D(3,0)$ | $B(2,0),\ C(2,0)$ |
| $A(0,1)$ | $B(0,2),\ B(1,1),\ C(0,2),\ C(1,1)$ |
| $D(1,2)$ | $C(0,2),\ C(1,1),\ B(0,2),\ B(1,1)$ |
| $B(0,2)$ | $A(0,1),\ D(0,3),\ D(1,2),\ C(0,2)$ |
| $C(0,2)$ | $A(0,1),\ D(0,3),\ D(1,2),\ B(0,2)$ |
| $D(0,3)$ | $B(0,2),\ C(0,2)$ |

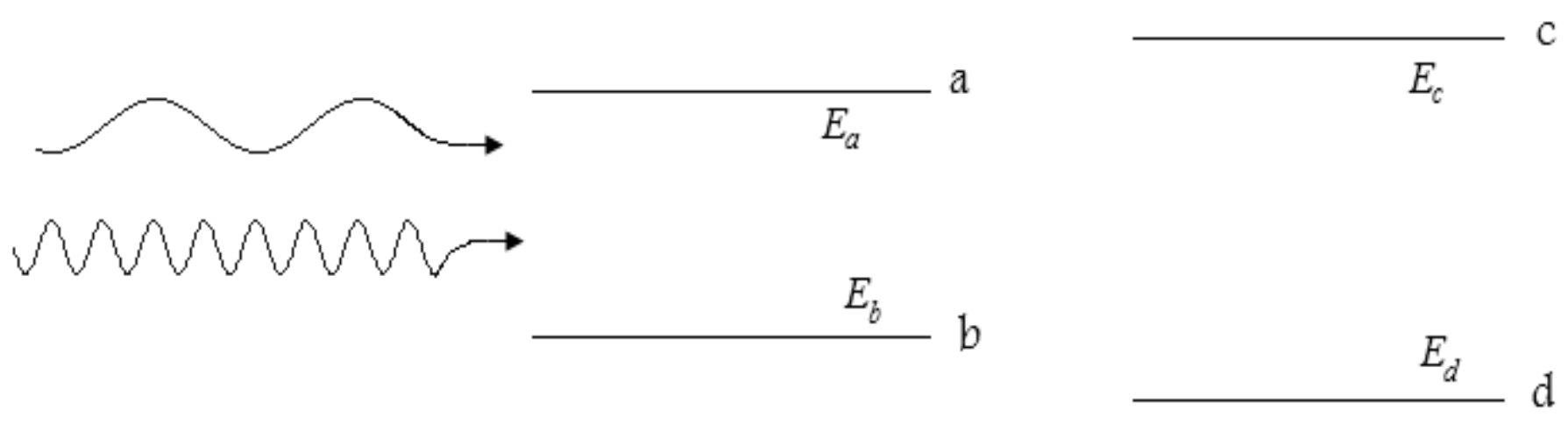

Figure 1.

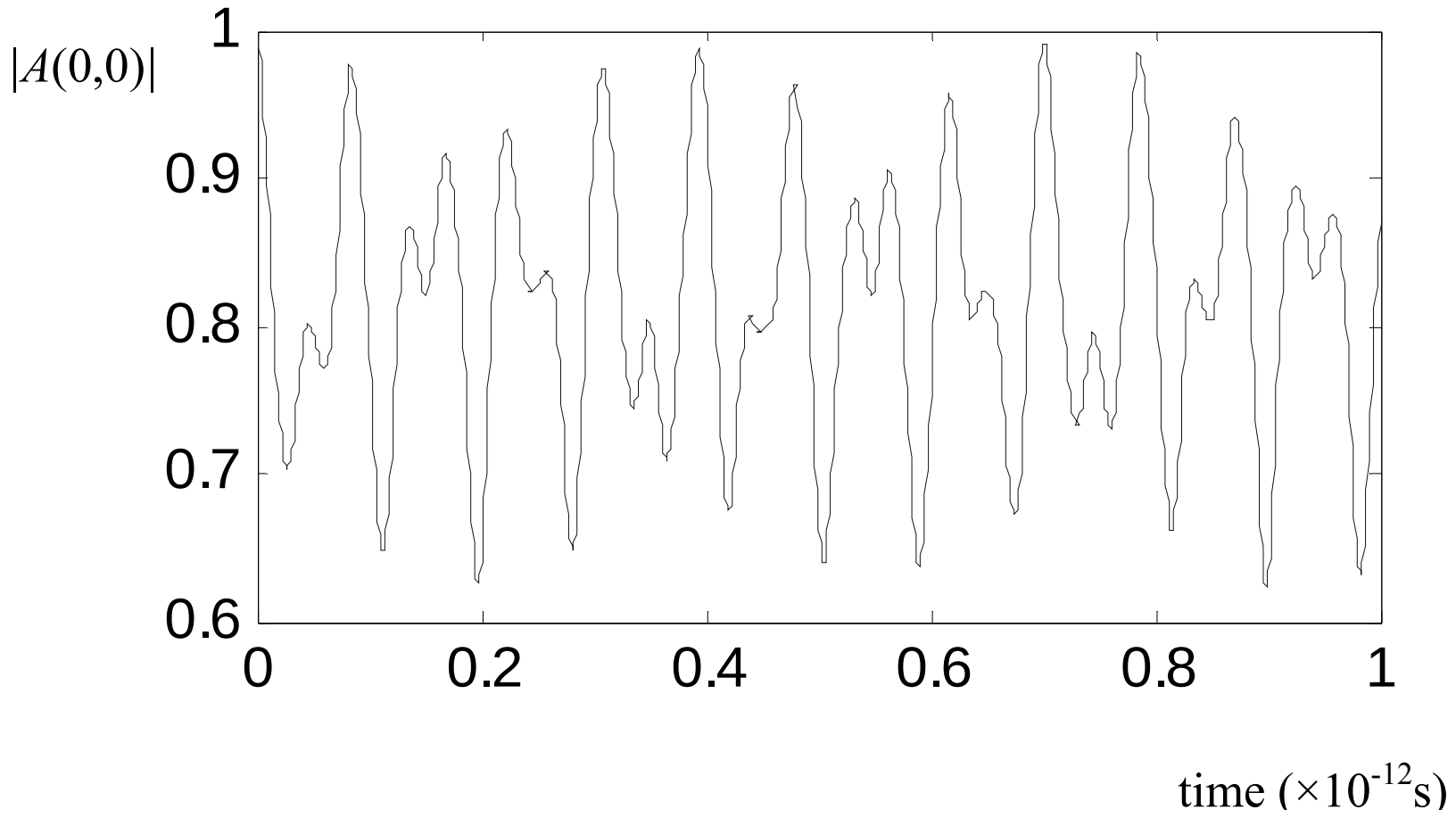

Figure 2.

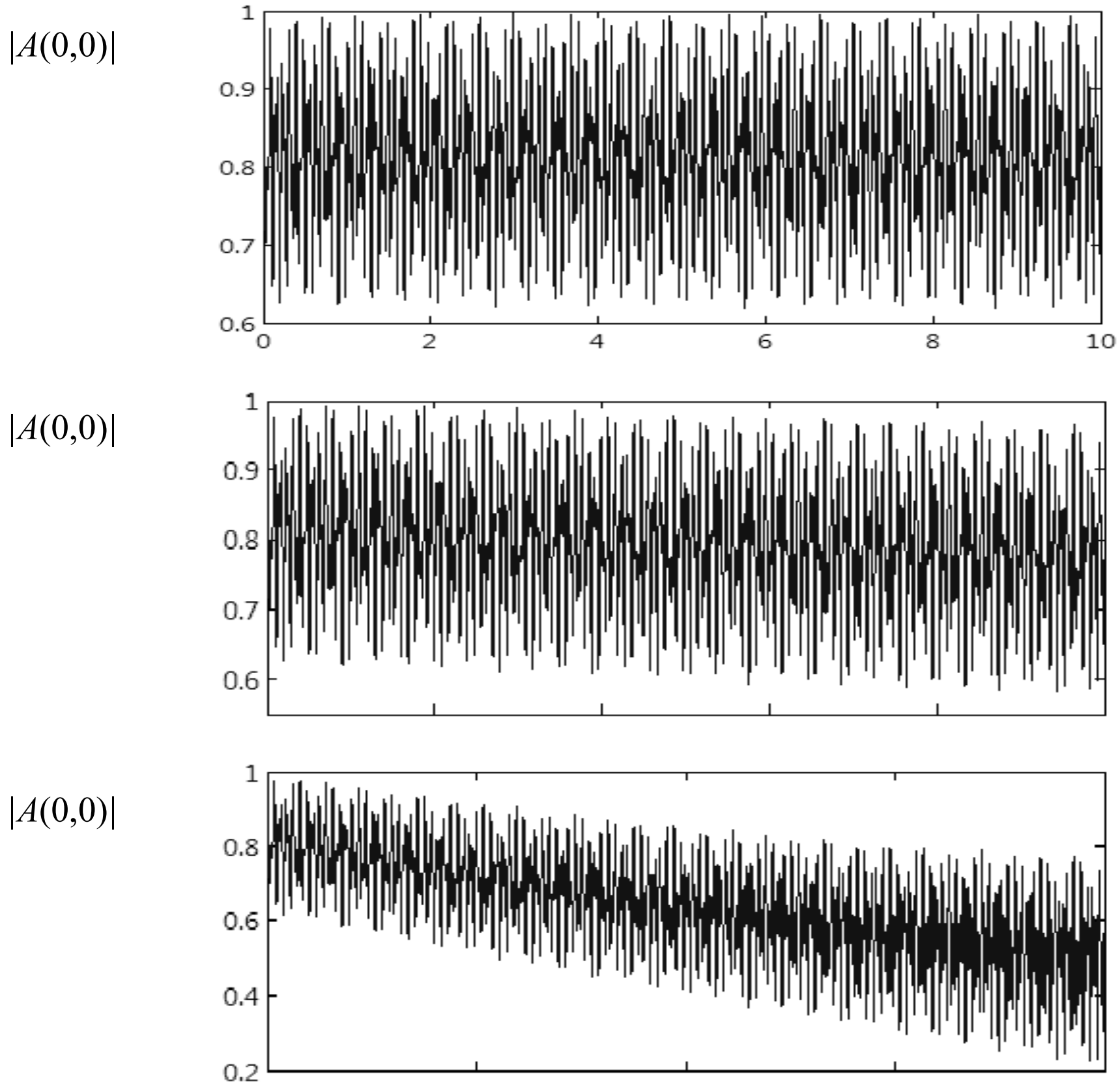

Figure 3.

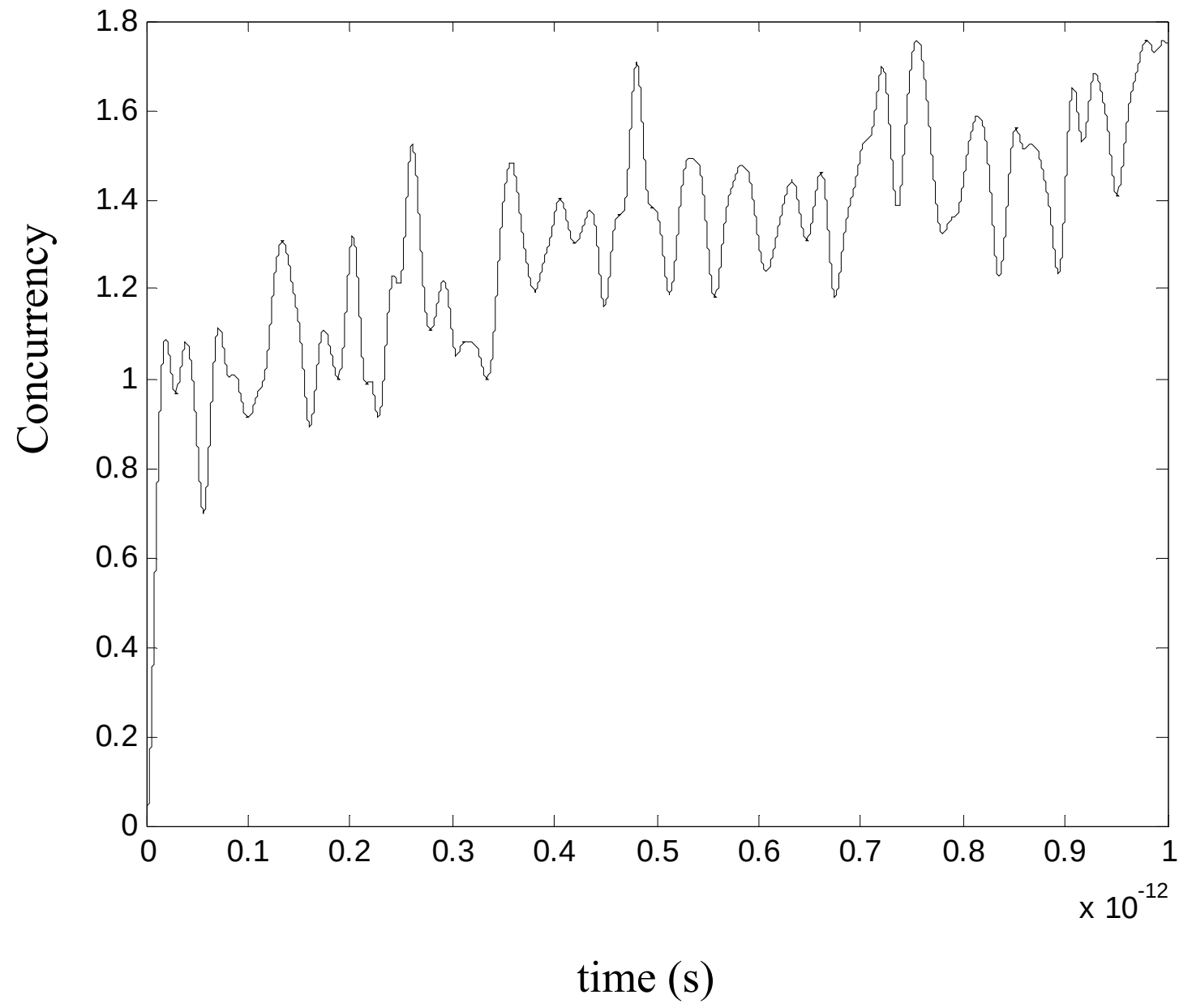

Figure 4.